\documentclass{article}

\PassOptionsToPackage{numbers, sort}{natbib}

\usepackage[preprint]{neurips_2026}
\newcommand{\eat}[1]{}

\usepackage{balance}
\usepackage{times}
\usepackage{latexsym}
\usepackage{amsfonts}
\usepackage{amsmath}
\usepackage{colortbl}
\usepackage{epsfig}
\usepackage{xspace}
\usepackage{graphicx}
\usepackage{comment}
\usepackage{booktabs}
\usepackage{multirow}
\usepackage{framed}
\usepackage{subcaption}

\usepackage{circledsteps}
\usepackage[utf8]{inputenc}
\usepackage{microtype}
\usepackage{inconsolata}
\usepackage[colorlinks=true, linkcolor=black, urlcolor=blue, citecolor=black]{hyperref}
\usepackage[most]{tcolorbox}
\usepackage{xcolor}

\usepackage[utf8]{inputenc}
\usepackage{CJKutf8}

\usepackage[T1]{fontenc}    
\usepackage{url}            
\usepackage{booktabs}       
\usepackage{amsfonts}       
\usepackage{nicefrac}       
\usepackage{microtype}      
\usepackage{xcolor}         

\usepackage[ruled,linesnumbered]{algorithm2e}
\usepackage{algpseudocode}

\algnewcommand\algorithmicinput{\textbf{Input:}}
\algnewcommand\algorithmicoutput{\textbf{Output:}}
\algnewcommand\Input{\item[\algorithmicinput]}
\algnewcommand\Output{\item[\algorithmicoutput]}

\algdef{SE}[DOWHILE]{Do}{doWhile}{\algorithmicdo}[1]{\algorithmicwhile\ #1}%

\definecolor{green}{RGB}{0,128,0}

\definecolor{yellow}{RGB}{255,200,18}

\newcommand{\stab}{\vspace{1.2ex}\noindent}

\newcommand{\bi}{\begin{itemize}}
\newcommand{\ei}{\end{itemize}}

\newcommand{\be}{\begin{enumerate}}
\newcommand{\ee}{\end{enumerate}}
\newcommand{\beqn}{\begin{eqnarray*}}
\newcommand{\eeqn}{\end{eqnarray*}}

\newcommand{\stitle}[1]{\stab\noindent{\bf #1}}

\usepackage{tabu}                      
\usepackage{booktabs}                  
\usepackage{lipsum}                    
\usepackage{mwe}                       

\usepackage{listings}
\usepackage{algorithm2e}

\usepackage{tikz}
\usetikzlibrary{arrows.meta,positioning,fit,calc,backgrounds}
\usepackage{pgfplots}
\pgfplotsset{compat=1.18}

\usepackage{enumitem}
\usepackage{natbib}
\usepackage{svg}
\usepackage{csquotes}
\usepackage{multirow}

\usepackage[normalem]{ulem}
\useunder{\uline}{\ul}{}

\usepackage{tabularx}
\usepackage{array}
\usepackage{booktabs}

\newcommand{\statcell}[2]{%
  \makecell[l]{#1 \\ \scriptsize #2}%
}

\newcommand{\sys}{\textbf{When Alpha Disappears}\xspace}

\setlist[itemize]{leftmargin=0.6cm, topsep=2pt, itemsep=1pt}  
\setlist[enumerate]{leftmargin=0.6cm, topsep=2pt, itemsep=1pt}  

\usepackage{circledsteps}
\pgfkeys{/csteps/fill color=black}
\pgfkeys{/csteps/inner color=white}

\definecolor{yy}{HTML}{f0c38e}
\definecolor{rr}{HTML}{f38181}

\usepackage{pifont}

\usepackage{float}
\usepackage{longtable}
\usepackage{makecell}

\title{\sys: A One-Switch Benchmark for Decision-Time Leakage in Financial Backtests}

\author{
\begin{tabular}{cccc}
Fan Zhang\textsuperscript{1} & Zhen Li\textsuperscript{1} & Sijia Peng\textsuperscript{2} & Yu Chen\textsuperscript{1}
\end{tabular}
\\[0.6em]
\textsuperscript{1}The University of Tokyo \\
\textsuperscript{2}HKUST (GZ) \\
\texttt{zhang-fan@g.ecc.u-tokyo.ac.jp}
}
\begin{document}

\maketitle

\begin{abstract}
  We introduce \sys, a paired evaluation benchmark for diagnosing decision-time leakage in financial machine-learning backtests. Rather than treating leakage as a binary property, the benchmark estimates protocol-induced inflation by toggling one evaluation convention at a time around a clean $t{+}1$-open reference, while holding the data panel, walk-forward split, model family, horizon, portfolio rule, and cost convention fixed. Across two daily-OHLCV equity panels, six model families, and yearly tests from 2016--2024, we find that inflation is highly selective: centered temporal features and same-day-open execution with post-open daily-bar information cause large and stable increases in both predictive and trading metrics, whereas global normalization, future-informed graph structure, and same-day-close execution are weak in most settings. The benchmark is diagnostic rather than a claim of tradable alpha, and is intended to make evaluation assumptions, failure modes, and protocol fragility directly measurable.
  \end{abstract}

\section{Introduction}

\begin{figure}[t]
  \centering
  \includegraphics[width=0.75\linewidth]{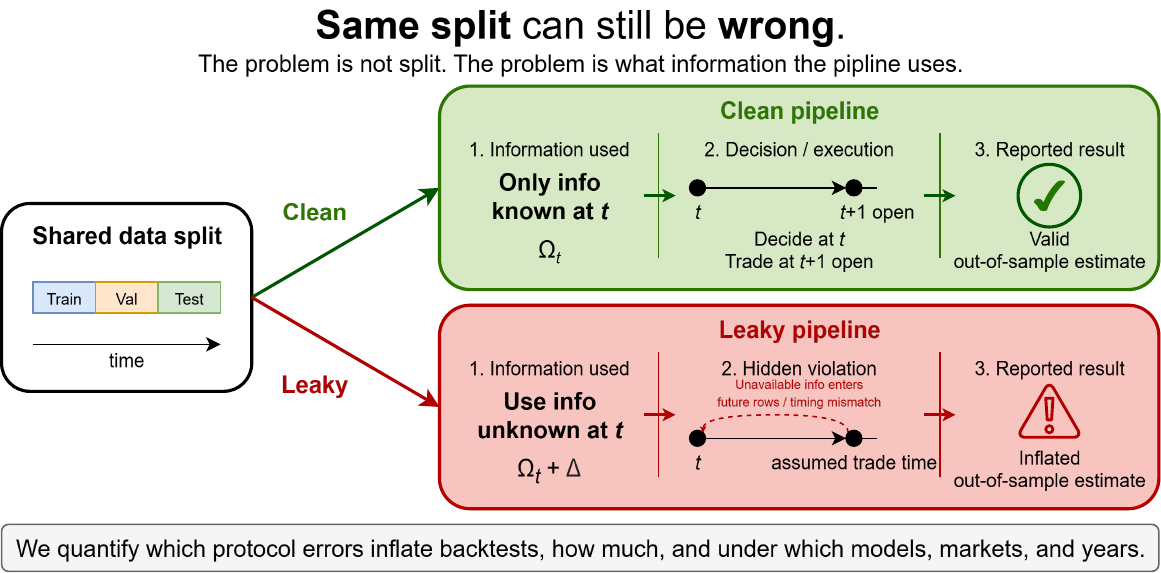}
  \caption{Same split, different information sets. Chronological order alone does not ensure that the signal information set and assumed fill time are decision-time valid.}
  \label{fig:motivation}
\end{figure}

\stitle{The same split can still be wrong.}
Historical backtesting is the dominant evaluation protocol in financial machine learning~\citep{de2018advances,bailey2014pseudo,koshiyama2019avoiding,gu2020empirical}. 
The standard safeguard is chronological separation: train on the past and evaluate on later data. 
Once this ordering is respected, the resulting Sharpe ratio or portfolio return is often treated as out-of-sample evidence of predictive skill~\citep{kaufman2012leakage,kapoor2023leakage,joeres2025data,sasse2023leakage}. 
However, chronological order alone does not define what information a model was allowed to use. 
Two backtests can use the same train/test split while relying on different information at the moment a trading decision is made. 
Figure~\ref{fig:motivation} illustrates this failure mode: the same split can support a clean pipeline that uses only available information, or a leaky pipeline that quietly changes the information set while still appearing out-of-sample.
In that case, a higher Sharpe no longer answers the intended scientific question: did the model forecast better, or did the evaluation pipeline quietly make the future easier to see?~\citep{bailey2017probability}

\stitle{Backtests must respect available information.}
A valid financial backtest requires more than temporal ordering. 
At each decision time $t$, the signal must be constructed using only information that would have been available before the decision is made. 
This constraint can be violated in many ways, including through rolling features, normalization scope, relation or graph estimation, target alignment, or execution assumptions~\citep{kaufman2012leakage,kapoor2023leakage,sasse2025overview,sasse2023leakage,bhand2026illusion,benhenda2026look,rosenblatt2024data}.
We refer to this requirement as \emph{decision-time semantics}. 
Violating it changes the effective information set of the backtest, producing results that appear out-of-sample but no longer correspond to a causal trading rule.

\stitle{The open problem is controlled attribution, not existence.}
Prior work has established that leakage can arise far beyond explicit train--test mixing, including through preprocessing, feature construction, validation design, and model selection~\citep{kaufman2012leakage,kapoor2023leakage,sasse2025overview,sasse2023leakage,bouke2024implications,karkar2026future,sarkar2025lookahead}. 
Finance-specific studies further show that temporally misaligned evaluation can distort economic conclusions~\citep{glasserman2023assessing,gao2025test,benhenda2026look,daniel2008look}, while live and post-cutoff benchmarks ask whether apparent performance survives outside historical replay~\citep{li2025timetravel,li2025profit}. 
These works establish that leakage matters, but they do not yet provide a controlled attribution framework. 
Within a fixed historical backtest, it remains unclear which protocol violations actually drive performance inflation, how large the marginal effect of each violation is, which model families are most vulnerable, and whether the effect persists across markets and years.

\stitle{Attribution requires one-switch benchmark design.}
This attribution problem is difficult because backtesting pipelines are highly entangled. 
Changing a feature construction rule may also change rankings, turnover, and transaction costs; modifying execution timing may affect both labels and realized returns; and relation-based features may interact with model architecture. 
As a result, differences between two backtests are not directly interpretable: a higher Sharpe may reflect a different model, a different portfolio rule, or a different information set. 
To isolate protocol-induced inflation, the evaluation must be turned into a one-switch benchmark comparison: hold fixed the data, walk-forward split, model family, portfolio rule, forecast horizon, and transaction-cost convention, and change only one protocol element at a time.

\stitle{Our benchmark: controlled attribution of protocol-induced inflation.}
We introduce \sys, a controlled benchmark for attributing backtest inflation to specific protocol violations in financial machine learning. 
Unlike model-comparison benchmarks, our primary object is not which forecaster performs best, but which protocol deviation changes reported performance under an otherwise matched evaluation grid. 
Starting from a decision-time-consistent $t\!+\!1$-open reference protocol, we instantiate this space with five controlled violations spanning temporal feature construction, normalization scope, graph construction, and execution alignment. 
Each variant modifies exactly one component of the evaluation pipeline. 
For each variant, we compute \emph{Leakage Gain}: the paired difference between the leaky run and the clean reference under identical data, model, portfolio rule, transaction-cost setting, horizon, and walk-forward split. 
This design isolates the marginal effect of individual protocol violations and makes evaluation fragility directly measurable.

Concretely, the clean reference fixes the decision-time semantics used for paired comparison: features are constructed using information available by the close of day \(t\), trades are executed at the \(t\!+\!1\) open, and labels are defined as open-to-open returns. 
Because each variant differs from this reference in exactly one protocol component, Leakage Gain measures the paired inflation induced by that deviation under the same trading setup.

Our contributions are as follows:
\begin{itemize}
  \item \textbf{Decision-time evaluation.} 
  We formulate financial backtesting around decision-time semantics and operationalize evaluation protocol fragility as a paired one-switch attribution benchmark over protocol violations, showing why chronological splitting alone is insufficient to define a valid out-of-sample evaluation.

  \item \textbf{Controlled attribution benchmark.} We propose \sys, a one-switch paired benchmark that isolates the marginal effect of individual protocol violations under fixed data, models, portfolio construction, horizons, costs, and walk-forward splits.

  \item \textbf{Finance-native leakage.} We identify execution-alignment violation as a distinct finance-native leakage mode: signals constructed using day-$t$ information can be paired with infeasible same-day execution assumptions.

  \item \textbf{Dual-market empirical evidence.} Across U.S. and China large-cap equity panels, two horizons, and six model families, we show that backtest inflation is selective, model-dependent, cross-market replicable, and stable across yearly splits.
\end{itemize}

\section{Related Work}
\label{sec:related-work}

\stitle{Financial ML and historical backtesting.}
Machine learning is now widely used in empirical asset pricing and return prediction, where models estimate conditional expected returns and are evaluated through historical backtests~\citep{gu2020empirical,de2018advances,bagnara2024asset}. 
These evaluations typically combine predictive metrics such as RankIC with portfolio-level metrics such as Sharpe ratios, cumulative returns, turnover, and drawdowns~\citep{de2018advances,khan2023performance,dessain2022machine}. 
A related literature studies backtest overfitting, data snooping, and the risk that repeated strategy search produces overstated out-of-sample performance~\citep{bailey2014pseudo,bailey2017probability,koshiyama2019avoiding}. 
Infrastructure work further supports reproducible financial machine learning through standardized data handling, model development, and backtesting workflows~\citep{yang2020qlib}. 
These works make it easier to compare models or strategies under a chosen evaluation protocol. 
Our work asks a different question: when does the protocol itself change the effective information set being evaluated?

\stitle{Leakage and evaluation fragility.}
Leakage is widely recognized as a central threat to empirical validity in machine learning~\citep{kaufman2012leakage,kapoor2023leakage,sasse2025overview,sasse2023leakage,nieto2026impact,sarkar2025lookahead}. 
Prior work shows that contamination can arise not only from explicit train--test mixing, but also from feature construction, preprocessing scope, model selection, validation design, and other pipeline decisions~\citep{kaufman2012leakage,sasse2025overview,bouke2024implications,joeres2025data,bayat2024pitfalls}. 
Such violations allow models to exploit information unavailable at prediction time, leading to overoptimistic estimates and failures to generalize~\citep{kapoor2023leakage,sasse2025overview,rosenblatt2024data,sarkar2025lookahead,karkar2026future}. 
In finance, analogous failures arise through rolling-window construction, normalization scope, target alignment, relation estimation, and execution semantics~\citep{glasserman2023assessing,gao2025test,benhenda2026look,daniel2008look}. 
Several recent studies further show that hidden leakage or spurious predictability can inflate financial results~\citep{bhand2026illusion,liu2026leakage,nikolopoulos2026spurious}. 
These works establish that leakage matters; \sys focuses on controlled attribution by placing multiple concrete violations into the same fixed data, model, horizon, portfolio, and cost grid, and measuring the paired inflation caused by one protocol switch at a time.

\stitle{Live evaluation and historical protocol diagnostics.}
Recent live and post-cutoff financial benchmarks raise a related but distinct concern: even chronologically clean historical evaluation can be misleading if pretrained models or agents have absorbed information about the evaluation period from their training corpus. 
Live benchmarking therefore asks whether apparent financial competence survives outside historical replay, where future data are not yet available to the model or evaluator~\citep{li2025timetravel,li2025profit,lopez2025memorization,roy2026memguard}. 
This perspective is complementary to ours. 
Live or post-cutoff evaluation tests whether performance survives prospective deployment; \sys asks which components of a historical backtesting pipeline inflate performance before deployment is even considered. 
In this sense, our benchmark treats the historical evaluation protocol itself as the object of diagnosis: it measures which protocol violations matter, how large their effects are, and whether those effects replicate across markets, model families, and yearly windows.

\section{The \sys Benchmark}
\label{sec:benchmark}

\subsection{Benchmark Design}
\label{sec:benchmark-design}

\begin{figure}[t]
  \centering
  \includegraphics[width=0.75\linewidth]{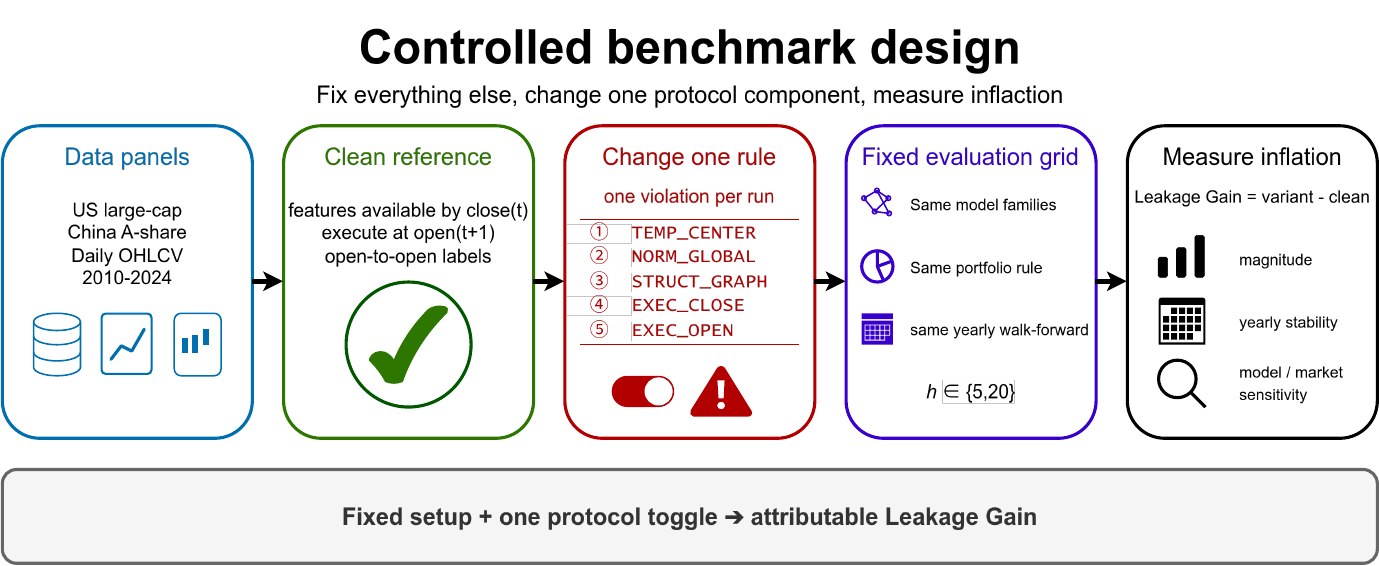}
  \caption{Controlled benchmark design. Within each pair, the data panel, split, model family, horizon, portfolio rule, cost setting, and metric are fixed; one protocol rule is toggled. Leakage Gain is the paired variant-minus-clean difference.}
  \label{fig:pipeline}
  \vspace{-5pt}
\end{figure}

\sys is a controlled benchmark for measuring how violations of decision-time semantics inflate financial backtests. 
The benchmark is organized around paired comparisons. 
For a fixed evaluation configuration, we run a clean decision-time reference protocol and compare it with five protocol variants: \texttt{TEMP\_CENTER}, \texttt{NORM\_GLOBAL}, \texttt{STRUCT\_GRAPH}, \texttt{EXEC\_CLOSE}, and \texttt{EXEC\_OPEN}. 
Each variant changes one protocol rule while holding the surrounding evaluation grid fixed.

Figure~\ref{fig:pipeline} summarizes the design. 
The key guarantee is that, within each pair, the market panel, yearly split, model family, forecast horizon, transaction-cost setting, and portfolio rule are shared. 
The only intended difference is the selected protocol switch.

\subsection{Decision-Time Reference}
\label{sec:clean-reference}

The clean reference defines the admissible information set. 
At the end of day \(t\), features may use only information available by the close of day \(t\). 
The model produces cross-sectional scores from this information set, the portfolio is entered at the open of day \(t+1\), and labels are defined as open-to-open returns over the forecast horizon. 
Thus, information availability, prediction time, execution time, and label construction are aligned under a \(t+1\)-open convention.

Portfolio construction is fixed across the clean and non-clean runs. 
On each evaluation date, stocks are ranked by the model score, the top decile is held long-only with equal weights, and transaction costs are applied using the same convention across protocol variants. 
This fixed portfolio rule prevents changes in reported performance from being attributed to a different trading rule.

Together, these conventions define the clean reference used in all paired Leakage Gain calculations. 
All protocol variants keep the surrounding data, model, portfolio, horizon, split, and cost settings fixed.

\subsection{Five One-Rule Protocol Violations}
\label{sec:protocol-violations}

Around the clean reference, we define five controlled violations. 
They instantiate different ways in which a backtest can preserve chronological train/test splits while changing the effective information set or the assumed execution time.

\stitle{\texttt{TEMP\_CENTER}.}
The clean reference uses trailing rolling operators. 
\texttt{TEMP\_CENTER} replaces them with centered rolling statistics, so a feature at decision time \(t\) can depend on observations after \(t\). 
This is a generic temporal-causality violation: the split remains chronological, but the feature construction no longer respects the decision-time information set.

\stitle{\texttt{NORM\_GLOBAL}.}
The clean reference fits normalization parameters within the training scope of each walk-forward window. 
\texttt{NORM\_GLOBAL} instead fits feature normalization using the full sample. 
This tests whether global preprocessing scope alone produces measurable inflation when the model, split, portfolio rule, and execution convention are otherwise unchanged.

\stitle{\texttt{STRUCT\_GRAPH}.}
The clean reference restricts relation or graph estimates to admissible past information within the corresponding walk-forward setting. 
\texttt{STRUCT\_GRAPH} replaces this with symmetric correlation graphs that use returns from both sides of the decision date. 
This violation targets leakage through estimated structure rather than through individual asset-level features.

\stitle{\texttt{EXEC\_CLOSE}.}
The clean reference forms signals after the close of day \(t\) and enters at the open of day \(t+1\). 
\texttt{EXEC\_CLOSE} instead pairs close-of-day signals with same-day close execution. 
This creates a timing inconsistency: information used to construct the signal is treated as if it were available for a fill at the same close.

\stitle{\texttt{EXEC\_OPEN}.}
\texttt{EXEC\_OPEN} is a finance-native execution-alignment violation.
It forms signals using the full OHLCV bar of day \(t\), including post-open fields such as high, low, close, and volume, while assuming execution at the open of the same day. 
This tests a finance-native leakage mode in which the backtest uses information that is unavailable at the assumed fill time.

Formal definitions of the clean information set, the five protocol switches, and their implementation-aligned execution conventions are provided in Appendix~\ref{app:formal-protocols}.

\subsection{Leakage Gain}
\label{sec:leakage-gain}

For metric \(q\), Leakage Gain is the paired difference between a protocol variant and the clean reference:
\[
\mathrm{LG}_{q}(v;m,a,h,c)
=
q(v;m,a,h,c)
-
q(\mathrm{CLEAN};m,a,h,c),
\]
where \(v\) is a protocol variant, \(m\) is a market, \(a\) is a model family, \(h\) is a forecast horizon, and \(c\) is a transaction-cost setting.

A positive Leakage Gain does not mean that the corresponding model is a better forecaster. 
It means that the reported metric increased after changing one protocol rule while keeping the rest of the evaluation grid fixed. 
Leakage Gain is therefore interpreted as protocol-induced inflation relative to the clean reference, rather than as absolute trading skill.

We compute Leakage Gain for both predictive and trading metrics. 
Predictive metrics include RankIC and AUC. 
Trading metrics include Sharpe ratios under proportional transaction costs, turnover, and maximum drawdown. 
In the main text, we emphasize \(\mathrm{LG\mbox{-}SR@5bps}\), because SR@5bps combines ranking quality, trading frequency, and transaction costs in a single trading-level measure~\citep{sharpe1998sharpe,lo2002statistics}. 
  Formal definitions of RankIC, AUC, \(\mathrm{SR@}c\mathrm{bps}\), \(\mathrm{LG\mbox{-}SR@}c\mathrm{bps}\), and the benchmark's per-day
  net return convention are given in Appendix~\ref{app:metric-definitions}.
For year-level stability analysis, the same paired difference is computed separately within each yearly walk-forward test window.

\paragraph{Scope.}
The benchmark is diagnostic rather than a claim of deployable trading alpha. 
It changes one daily-OHLCV protocol rule at a time under fixed universe, split, model, portfolio, horizon, and cost conventions. 
It does not model full point-in-time universe maintenance, capacity, liquidity, market impact, intraday execution, or interacting leakage sources; implementation details are provided in Appendix~\ref{app:implementation}.

\section{Experiments and Results}
\label{sec:experiments}

\stitle{Research questions.}
We organize the experiments around four questions.
RQ1: Which protocol violations inflate backtest performance?
RQ2: Is inflation selective, or does every relaxed protocol mechanically improve results?
RQ3: Does leakage susceptibility depend on model family or market environment?
RQ4: Are the dominant effects stable across yearly walk-forward windows?

\subsection{Setting}
\label{sec:setting}

We evaluate two daily OHLCV equity panels with shared protocol semantics. 
The U.S. panel is derived from S\&P 500 constituents and covers 2010-01-04 to 2024-12-30, with 439 retained tickers and daily cross-sections ranging from 422 to 439 stocks. 
The China panel is derived from CSI 300 constituents and covers 2010-01-04 to 2024-12-31, with 186 retained tickers and daily cross-sections ranging from 135 to 186 stocks. 
Both panels use yearly walk-forward tests over 2016--2024.

We evaluate six model families: Momentum, Ridge, LightGBM, a graph-informed MLP, an LSTM, and a Transformer. 
All models are run on both markets, horizons \(h\in\{5,20\}\), and the clean plus five protocol variants from Section~\ref{sec:protocol-violations}. 
The portfolio and metrics follow Section~\ref{sec:leakage-gain}: top-decile long-only equal-weight portfolios, proportional costs at 0/5/10 bps, and LG-SR@5bps as the main scalar summary. 
Full panel, split, model, protocol-switch, and metric details are provided in Appendices~\ref{app:formal-protocols} and~\ref{app:implementation}.

\subsection{Clean Reference Performance}
\label{sec:clean-baselines}

Before measuring inflation, we first establish the scale of clean reference performance. 
Clean SR@5bps is intentionally modest. 
On the U.S. panel, it ranges from \(-0.20\) to \(0.68\) at \(h=5\) and from \(0.08\) to \(0.79\) at \(h=20\) across the six model families. 
On the China panel, the corresponding ranges are \(-0.01\) to \(0.93\) and \(0.44\) to \(0.93\).

Ridge and Momentum serve different diagnostic roles. 
Ridge is a trainable model whose scores depend directly on the feature construction pipeline. 
It is therefore sensitive to temporal and preprocessing violations when those violations alter the information contained in the input features. 
Momentum, by contrast, uses a fixed \texttt{ret\_20} score and is nearly invariant to feature-construction toggles such as \texttt{TEMP\_CENTER}. 
However, it can still react strongly to execution-time violations because the same ranking rule is evaluated under a different trading-time assumption. 
This separation helps distinguish feature-level leakage from execution-alignment leakage.

\subsection{Main Results: Two Violations Dominate}
\label{sec:main-results}

Table~\ref{tab:main-results} is the main benchmark table.
Each row reports the clean SR@5bps and the incremental \(\mathrm{LG\mbox{-}SR@5bps}\) under each protocol violation for one market--model--horizon configuration.
The pattern is immediate: \texttt{TEMP\_CENTER} and \texttt{EXEC\_OPEN} dominate across markets and model families, while \texttt{NORM\_GLOBAL}, \texttt{STRUCT\_GRAPH}, and \texttt{EXEC\_CLOSE} are weak or near-zero in most settings.

These magnitudes should not be read as deployable alpha. For the dominant violations, especially \texttt{TEMP\_CENTER} and \texttt{EXEC\_OPEN}, the benchmark deliberately constructs infeasible, oracle-like protocols that relax the decision-time information set. Large \(\mathrm{LG\mbox{-}SR@5bps}\) values are therefore the expected diagnostic signal of protocol-induced inflation, rather than evidence of a realistic trading edge.

\begin{table*}[t]
  \centering
  \small
  \caption{Main benchmark table. Each row shows the clean Sharpe at 5 bps and the corresponding Leakage Gain in Sharpe under each
  protocol violation. \texttt{TEMP\_CENTER} and \texttt{EXEC\_OPEN} dominate, while the other violations are weak or near-zero. Large
  gains under the dominant violations reflect diagnostic inflation under infeasible protocols rather than deployable alpha, and therefore are not directly comparable to deployable performance.}
  \label{tab:main-results}
  \resizebox{\textwidth}{!}{%
  \begin{tabular}{llcrrrrrr}
    \toprule
    Market & Model & $h$ & Clean & \textbf{TEMP\_CENTER} & \textbf{EXEC\_OPEN} & NORM\_GLOBAL & STRUCT\_GRAPH & EXEC\_CLOSE \\
    \midrule
    \multirow{12}{*}{\shortstack{US\\S\&P 500}} & \multirow{2}{*}{Momentum} & 5 & 0.49 & -0.00 & \textbf{5.41} & 0.00 & 0.00 & 0.03 \\
     &  & 20 & 0.53 & -0.00 & \textbf{5.40} & 0.00 & 0.00 & 0.02 \\
     & \multirow{2}{*}{Ridge} & 5 & 0.68 & 19.43 & \textbf{21.65} & 0.00 & -0.05 & -0.03 \\
     &  & 20 & 0.79 & 17.43 & \textbf{17.69} & 0.00 & -0.03 & -0.05 \\
     & \multirow{2}{*}{LightGBM} & 5 & 0.57 & 20.11 & \textbf{22.83} & 0.01 & 0.04 & -0.03 \\
     &  & 20 & 0.78 & \textbf{17.00} & 16.73 & 0.04 & -0.02 & -0.07 \\
     & \multirow{2}{*}{Graph} & 5 & 0.33 & 18.83 & \textbf{21.06} & 0.04 & 0.06 & -0.01 \\
     &  & 20 & 0.51 & \textbf{15.97} & 14.66 & 0.13 & 0.32 & 0.02 \\
     & \multirow{2}{*}{LSTM} & 5 & 0.25 & \textbf{4.67} & 4.26 & 0.01 & 0.02 & 0.21 \\
     &  & 20 & 0.33 & \textbf{4.94} & 4.57 & 0.02 & 0.04 & -0.08 \\
     & \multirow{2}{*}{Transformer} & 5 & -0.20 & \textbf{9.19} & 6.09 & 0.38 & 0.26 & 0.41 \\
     &  & 20 & 0.08 & \textbf{6.11} & 6.08 & 0.11 & 0.06 & 0.15 \\
    \midrule
    \multirow{12}{*}{\shortstack{China\\CSI 300}} & \multirow{2}{*}{Momentum} & 5 & 0.44 & 0.00 & \textbf{6.29} & 0.00 & 0.00 & 0.11 \\
     &  & 20 & 0.44 & 0.00 & \textbf{6.29} & 0.00 & 0.00 & 0.11 \\
     & \multirow{2}{*}{Ridge} & 5 & 0.84 & 21.70 & \textbf{26.16} & 0.00 & -0.05 & -0.16 \\
     &  & 20 & 0.93 & 20.74 & \textbf{22.18} & 0.00 & -0.17 & 0.11 \\
     & \multirow{2}{*}{LightGBM} & 5 & 0.93 & 21.78 & \textbf{26.04} & 0.00 & 0.37 & 0.48 \\
     &  & 20 & 0.78 & 20.55 & \textbf{21.89} & 0.03 & 0.31 & 0.22 \\
     & \multirow{2}{*}{Graph} & 5 & 0.62 & 20.89 & \textbf{23.82} & -0.02 & 0.15 & -0.15 \\
     &  & 20 & 0.57 & 19.44 & \textbf{19.70} & 0.06 & 0.24 & 0.24 \\
     & \multirow{2}{*}{LSTM} & 5 & 0.43 & \textbf{10.31} & 7.64 & -0.14 & -0.23 & -0.07 \\
     &  & 20 & 0.56 & 6.83 & \textbf{8.01} & -0.07 & -0.18 & 0.01 \\
     & \multirow{2}{*}{Transformer} & 5 & -0.01 & \textbf{12.40} & 10.79 & 0.08 & 0.37 & 0.08 \\
     &  & 20 & 0.58 & \textbf{13.22} & 9.62 & -0.06 & -0.19 & -0.31 \\
    \bottomrule
  \end{tabular}%
  }
\end{table*}

For Ridge, both dominant violations create large inflation on both markets.
At \(h=5\), \texttt{TEMP\_CENTER} increases SR@5bps by \(19.43\) on the U.S. panel and \(21.70\) on the China panel, while \texttt{EXEC\_OPEN} increases it by \(21.65\) and \(26.16\), respectively.
The same conclusion holds at \(h=20\), where both dominant violations remain large in both markets.
This shows that the effect is not confined to one forecast horizon or one regional market.

The fixed momentum baseline sharpens the interpretation.
Momentum is essentially unaffected by \texttt{TEMP\_CENTER}, as expected, because its score does not depend on the centered rolling features.
However, it reacts strongly to \texttt{EXEC\_OPEN}, with \(\mathrm{LG\mbox{-}SR@5bps}\) around \(5\)--\(6\) on both markets and horizons.
This confirms that \texttt{EXEC\_OPEN} is not merely another version of temporal feature leakage.
It is a distinct execution-alignment violation: the backtest uses information from the day-\(t\) bar while assuming a same-day-open fill.

The non-linear model families show the same ordering.
LightGBM and the graph-informed model are the most sensitive, with \texttt{TEMP\_CENTER} and \texttt{EXEC\_OPEN} often producing \(\mathrm{LG\mbox{-}SR@5bps}\) above \(15\) and sometimes above \(20\).
The sequence models also inflate materially, although with smaller magnitudes on the U.S. panel.
Across the full benchmark, the qualitative conclusion is stable: the largest gains come from a generic temporal-causality leak and an execution-alignment violation, not from every possible relaxation of the protocol.

\stitle{RQ1.}
The protocol violations that materially inflate backtest performance are selective rather than universal.
Across two markets, two horizons, and six model families, \texttt{TEMP\_CENTER} and \texttt{EXEC\_OPEN} dominate, while \texttt{NORM\_GLOBAL}, \texttt{STRUCT\_GRAPH}, and \texttt{EXEC\_CLOSE} are weak or near-zero in most settings.

\subsection{Negative Controls and Semantic Sanity Checks}
\label{sec:sanity-checks}

The benchmark would be much less convincing if every relaxed protocol mechanically improved results.
That is not what we observe.
Across the full model-family extension, \texttt{NORM\_GLOBAL}, \texttt{STRUCT\_GRAPH}, and \texttt{EXEC\_CLOSE} remain weak and often near-zero after transaction costs.
Averaged across LightGBM, Graph, LSTM, and Transformer at both horizons, the mean \(\mathrm{LG\mbox{-}SR@5bps}\) on the U.S. panel is only \(0.07\) for \texttt{EXEC\_CLOSE}, \(0.09\) for \texttt{NORM\_GLOBAL}, and \(0.10\) for \texttt{STRUCT\_GRAPH}.
On the China panel, the corresponding means are \(0.06\), \(-0.01\), and \(0.11\).
These negative controls show that the benchmark is not simply assigning larger numbers to every non-clean configuration.

This ordering is also not tied to the 5 bps operating point.
In an additional Ridge \(h=5\) cost-sensitivity check over \(0/5/10/25/50\) bps, the two dominant violations remain far above the clean protocol on both markets throughout the full range.
Appendix Figure~\ref{fig:appendix-cost-sensitivity} summarizes this result.

The dominant violations also produce coherent changes in trading-side quantities.
For Ridge at \(h=5\), turnover rises from \(0.746\) under the clean protocol to \(1.171\) under \texttt{TEMP\_CENTER} and \(1.697\) under \texttt{EXEC\_OPEN}.
At \(h=20\), it rises from \(0.399\) to \(1.064\) and \(1.444\).
These changes are consistent with more reactive, information-privileged rankings rather than a reporting artifact in a single metric.

\begin{figure}[t]
  \centering
  \includegraphics[width=0.80\linewidth]{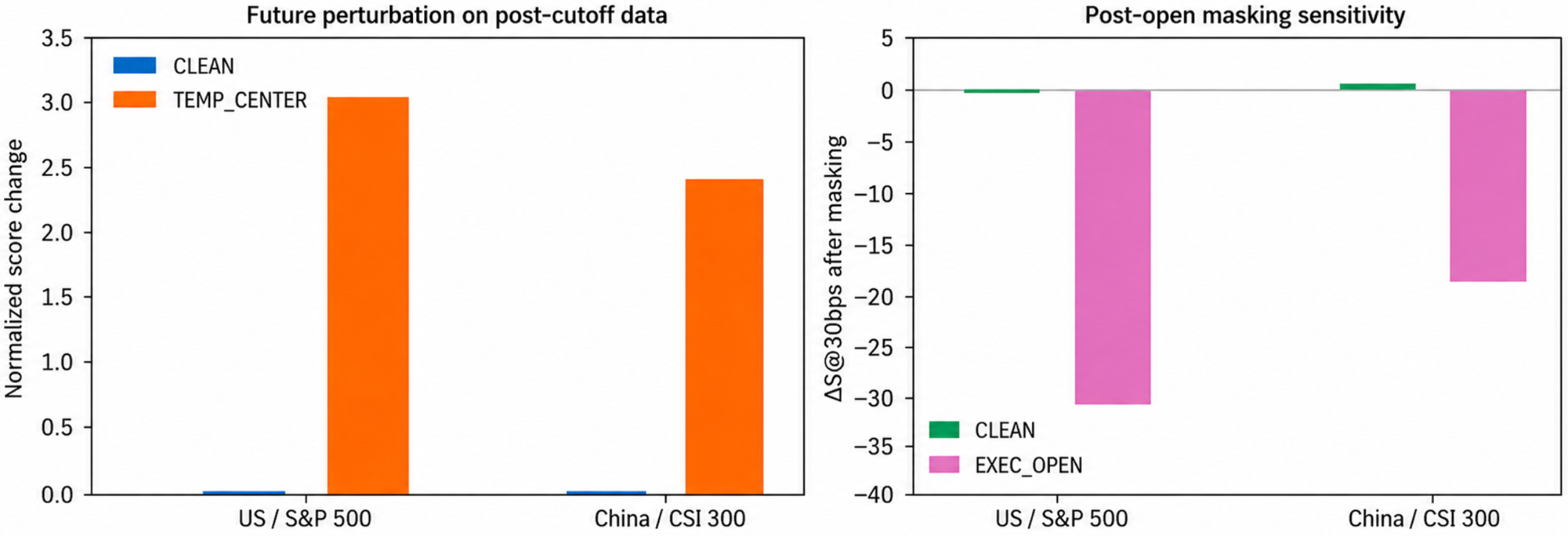}
  \caption{Semantic sanity checks on the 2024 Ridge window. Future-suffix perturbations affect \texttt{TEMP\_CENTER} but not \texttt{CLEAN}; post-open masking collapses \texttt{EXEC\_OPEN} but barely affects \texttt{CLEAN}.}
  \label{fig:sanity-interventions}
\end{figure}

Finally, we run two intervention-style sanity checks on the 2024 test window using the core-feature Ridge baseline.
First, we perturb the post-cutoff future suffix while holding the admissible history and open-to-open labels fixed.
This leaves the clean protocol unchanged but moves \texttt{TEMP\_CENTER} materially, with normalized score changes of \(3.06\) on the U.S. panel and \(2.41\) on the China panel.
Second, we mask same-day post-open fields by replacing high, low, close, and volume with open-time admissible surrogates.
Under this masking, the clean protocol changes only mildly, whereas \texttt{EXEC\_OPEN} collapses: \(\Delta\)SR@5bps is \(-35.35\) on the U.S. panel and \(-20.43\) on the China panel.
Figure~\ref{fig:sanity-interventions} summarizes these interventions.

\stitle{RQ2.}
Inflation is selective and semantically grounded rather than a mechanical reward for every relaxed protocol.
The largest gains depend on the specific inadmissible information introduced by the corresponding violation.

\subsection{Model-Family and Market Heterogeneity}
\label{sec:model-market-heterogeneity}
\begin{figure*}[t]
  \centering
  \includegraphics[width=0.95\textwidth]{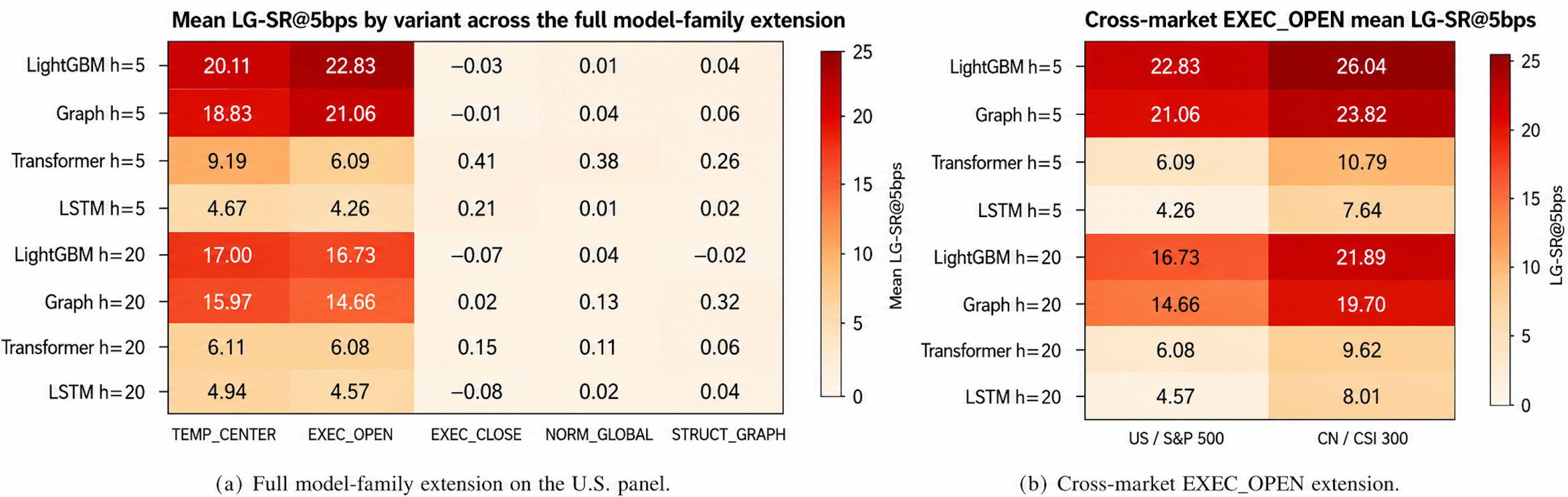}
  \caption{Model and market heterogeneity. Left: U.S. mean LG-SR@5bps by protocol variant for extended models. Right: cross-market comparison of \texttt{EXEC\_OPEN} gains.}
  \label{fig:model-family}
\end{figure*}
We then evaluate whether leakage susceptibility depends on model family and market environment.
Figure~\ref{fig:model-family} shows the full model-family extension across LightGBM, Graph, LSTM, and Transformer.
The dominant violations remain inflationary across all four families, but the magnitude differs substantially.

For \texttt{TEMP\_CENTER}, the average \(\mathrm{LG\mbox{-}SR@5bps}\) across the four non-linear families and two horizons is \(12.10\) on the U.S. panel and \(15.68\) on the China panel.
For \texttt{EXEC\_OPEN}, the corresponding averages are \(12.04\) and \(15.94\).
Tree-based and graph-informed models are the most vulnerable.
At \(h=5\), LightGBM and Graph reach \(\mathrm{LG\mbox{-}SR@5bps}\) of \(22.83\) and \(21.06\) under \texttt{EXEC\_OPEN} on the U.S. panel, and \(26.04\) and \(23.82\) on the China panel.
Sequence models are less extreme but still materially affected; for example, under \texttt{EXEC\_OPEN}, Transformer reaches \(10.79\) on the China panel at \(h=5\), and LSTM reaches \(8.01\) at \(h=20\).

The market comparison is also informative.
The China panel is at least as sensitive as the U.S. panel under the dominant violations and often more sensitive, especially for sequence models.
This suggests that protocol fragility is not only a property of model architecture; it also interacts with the market environment and data-generating process.

\stitle{RQ3.}
Leakage susceptibility is model- and market-dependent.
Tree and graph models are the most sensitive, but sequence models also inflate materially, especially on the China panel.

\subsection{Year-Level Stability}
\label{sec:year-stability}

\begin{table*}[t]
\centering
\scriptsize
\caption{Year-level stability of the two dominant violations. Each cell reports mean yearly $\mathrm{LG\mbox{-}SR@5bps}$ with the number of positive years in parentheses. Full confidence intervals and Wilcoxon tests are reported in Appendix~\ref{app:year-stability}.}
\label{tab:year-stability-main}
\setlength{\tabcolsep}{3.2pt}
\renewcommand{\arraystretch}{1.08}

\resizebox{\textwidth}{!}{
\begin{tabular}{lcccccccc}
\toprule
\multirow{2}{*}{Model}
& \multicolumn{4}{c}{\textbf{US / S\&P 500}}
& \multicolumn{4}{c}{\textbf{China / CSI 300}} \\
\cmidrule(lr){2-5} \cmidrule(lr){6-9}
& \multicolumn{2}{c}{\textbf{TEMP\_CENTER}}
& \multicolumn{2}{c}{\textbf{EXEC\_OPEN}}
& \multicolumn{2}{c}{\textbf{TEMP\_CENTER}}
& \multicolumn{2}{c}{\textbf{EXEC\_OPEN}} \\
\cmidrule(lr){2-3} \cmidrule(lr){4-5}
\cmidrule(lr){6-7} \cmidrule(lr){8-9}
& $h=5$ & $h=20$ & $h=5$ & $h=20$
& $h=5$ & $h=20$ & $h=5$ & $h=20$ \\
\midrule

Momentum
& -0.00 (4/9) & -0.00 (4/9)
& \textbf{6.57 (9/9)} & \textbf{6.58 (9/9)}
& 0.00 (0/9) & 0.00 (0/9)
& \textbf{6.58 (9/9)} & \textbf{6.58 (9/9)} \\

Ridge
& 25.20 (9/9) & 22.45 (9/9)
& \textbf{29.10 (9/9)} & \textbf{23.50 (9/9)}
& 23.18 (9/9) & 22.17 (9/9)
& \textbf{28.33 (9/9)} & \textbf{23.80 (9/9)} \\

LightGBM
& 25.24 (9/9) & \textbf{21.88 (9/9)}
& \textbf{29.77 (9/9)} & 21.78 (9/9)
& 23.18 (9/9) & 22.13 (9/9)
& \textbf{28.35 (9/9)} & \textbf{24.18 (9/9)} \\

Graph
& 23.96 (9/9) & \textbf{20.58 (9/9)}
& \textbf{26.74 (9/9)} & 20.18 (9/9)
& 22.40 (9/9) & 20.88 (9/9)
& \textbf{25.92 (9/9)} & \textbf{21.24 (9/9)} \\

LSTM
& \textbf{6.22 (9/9)} & \textbf{5.93 (9/9)}
& 5.38 (8/9) & 5.65 (9/9)
& \textbf{11.52 (9/9)} & 7.86 (9/9)
& 8.47 (9/9) & \textbf{8.83 (9/9)} \\

Transformer
& \textbf{11.82 (9/9)} & \textbf{7.29 (9/9)}
& 7.29 (9/9) & 7.13 (9/9)
& \textbf{13.19 (9/9)} & \textbf{14.22 (9/9)}
& 11.56 (9/9) & 11.09 (9/9) \\

\bottomrule
\end{tabular}
}

\vspace{2pt}
\begin{minipage}{0.96\textwidth}
\footnotesize
\emph{Note.} Boldface marks the larger violation within each market--model--horizon setting. Parentheses report the number of years with positive Leakage Gain.
\end{minipage}
\end{table*}

Table~\ref{tab:year-stability-main} summarizes yearly stability using the mean yearly Leakage Gain and the number of positive years. Full confidence intervals and paired Wilcoxon tests are reported in Appendix~\ref{app:year-stability}. A pooled result can be misleading if it is driven by a few favorable windows.
We therefore compute yearly paired Leakage Gains relative to the clean benchmark.
For the two dominant violations, we report the mean yearly \(\mathrm{LG\mbox{-}SR@5bps}\), bootstrap 95\% confidence intervals, the number of positive years, and a one-sided paired Wilcoxon signed-rank test.
Full yearly statistics for the dominant violations are also reported in Appendix~\ref{app:year-stability}.

Figure~\ref{fig:yearly-exec-open} illustrates the yearly pattern for Ridge under \texttt{EXEC\_OPEN}.
The gains are positive in every yearly window on both markets and both horizons.
The full model-family extension shows the same qualitative pattern: the dominant gains are broadly positive rather than concentrated in a small number of favorable years.

\begin{figure}[t]
  \centering
  \includegraphics[width=0.85\linewidth]{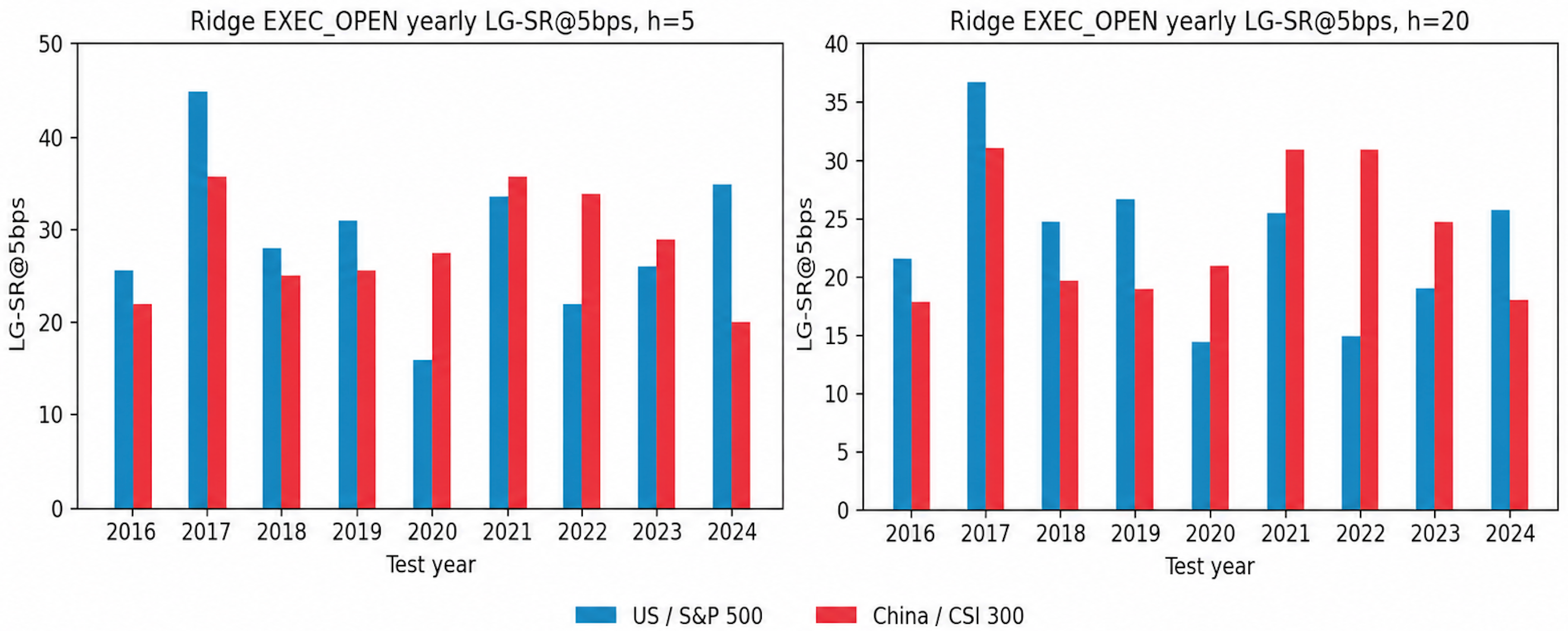}
  \caption{Yearly \(\mathrm{LG\mbox{-}SR@5bps}\) for Ridge under \texttt{EXEC\_OPEN}.
  The positive effect is not concentrated in a small number of windows.}
  \label{fig:yearly-exec-open}
\end{figure}

\stitle{RQ4.}
The dominant inflation sources are year-level stable.
They are not artifacts of a small number of unusually favorable test windows.




\section{Conclusion}

  We presented \sys, a controlled benchmark for measuring backtest inflation under protocol violations. Across paired comparisons around
  a clean decision-time reference, we show that strong apparent out-of-sample performance can arise even when chronological splitting is
  preserved. Two violations dominate: \texttt{TEMP\_CENTER}, a generic temporal leak, and \texttt{EXEC\_OPEN}, a finance-native
  execution leak.

  The main implication is that chronological splitting alone is not a sufficient guarantee of clean evaluation in financial machine
  learning. Decision-time semantics and protocol fragility should therefore be treated as explicit benchmark objects. Future work can
  extend the same paired design to additional markets, model classes, and interacting violations.

\section*{Limitation}
\label{sec:limitation}


  This work is a diagnostic benchmark under a fixed daily-bar evaluation setup. It studies two large-cap equity markets, long-only top-decile portfolios, proportional transaction costs, and one protocol violation at a time. It does not model interacting leakage sources, full point-in-time universe maintenance, capacity, market impact, or intraday execution. These boundaries limit claims about absolute tradability, but not the central paired result: specific protocol deviations can materially inflate reported backtest performance under matched evaluation settings.

\clearpage
\newpage

\bibliographystyle{plainnat}
\bibliography{refs}

\clearpage
\newpage

\appendix
\appendix

\section{Broader Impacts}
\label{sec:impacts}

This work may have positive impact by improving the reliability and transparency of financial model evaluation. By showing how specific protocol violations inflate backtest performance, the benchmark helps researchers and practitioners distinguish genuine predictive evidence from artifacts of the evaluation pipeline. More explicit decision-time semantics and execution assumptions can support more credible empirical finance research and reduce the risk of deploying strategies based on misleading historical results.

The work also has potential negative or dual-use implications. If the findings are misunderstood or ignored, inflated backtests may encourage overconfident deployment of trading strategies, leading to financial losses; in aggregate or under leveraged deployment, such failures may contribute to broader market risk. Conversely, knowledge of high-impact protocol violations could be misused to construct evaluations that exaggerate model performance.

We mitigate these risks through the benchmark design: clean reference protocols, controlled attribution, transparent reporting of protocol assumptions, and independent validation under leakage-resistant evaluation procedures.

\section{Ethical Statement}
\label{sec:ethics}

We use public or properly licensed market data and software components, and we do not redistribute restricted raw financial data when licenses do not permit redistribution. 
The benchmark does not use personally identifiable information, individual investor records, or private trading data.

The artifact is a research benchmark for protocol-fragility diagnosis; it is not an autonomous trading system or investment recommendation. 
Any downstream use should include domain-expert review, independent leakage-resistant validation, appropriate risk controls, and compliance with applicable laws, regulations, and data licenses.

\section{Benchmark Formalization}
\label{app:formal-protocols}

\subsection{Formal Protocol Definitions}
\label{app:formal-switches}

This benchmark is designed around \emph{one-rule attribution}: each leaky variant changes one protocol component while keeping the data source, walk-forward split, model family, and portfolio rule fixed.
To make that claim precise, we formalize the clean reference and the five protocol switches below.

\paragraph{Notation.}
For asset \(i\) and trading day \(t\), let
\[
b_{i,t} = (O_{i,t}, H_{i,t}, L_{i,t}, C_{i,t}, V_{i,t})
\]
denote the daily OHLCV bar, and let \(h \in \{5,20\}\) be the forecasting horizon.
The benchmark constructs a feature vector \(x_{i,t}\), a regression target \(y_{i,t}\), and a one-day realized trade return \(r^{\mathrm{trade}}_{i,t}\) used by the long-only backtest.
The portfolio always ranks scores cross-sectionally on day \(t\) and holds the top decile with equal weights.

\paragraph{Clean reference.}
In the clean setting, signals are formed after the close of day \(t\), execution occurs at the open of day \(t+1\), and the label is open-to-open:
\begin{align}
y^{\mathrm{clean}}_{i,t}
&= \log \frac{O_{i,t+h+1}}{O_{i,t+1}}, \\
r^{\mathrm{trade,clean}}_{i,t}
&= \frac{O_{i,t+2}}{O_{i,t+1}} - 1 .
\end{align}
Feature standardization parameters are fit on the training split only, and graph/relation features are estimated from admissible trailing information only.

\paragraph{Clean graph construction.}
Let \(m(t)\) be the calendar month containing day \(t\), and let \(\tau_m\) be the first trading day of month \(m\).
For each month \(m\), the clean relation graph is estimated once using the trailing return window
\[
\mathcal{W}^{\mathrm{trail}}_m
=
\{ s : s < \tau_m \}^{\text{last }252},
\]
i.e., the last \(252\) trading days strictly before the monthly anchor \(\tau_m\).
From this window we compute the correlation matrix of asset close-to-close log returns, take absolute values, set the diagonal to zero, keep the top-\(k\) peers for each asset, and normalize each nonzero row to sum to one.
Denoting the resulting weight matrix by \(W^{(m),\mathrm{trail}}\), the relation features used on any day \(t \in m\) take the form
\[
\mathrm{nbr\_feat}_{i,t}
=
\sum_j W^{(m),\mathrm{trail}}_{ij} \, \mathrm{feat}_{j,t},
\]
for \(\mathrm{feat} \in \{\mathrm{ret}_5,\mathrm{ret}_{20},\mathrm{vol\_ratio}_{20},\mathrm{hl\_range\_5\_mean}\}\).

\subsubsection{Five One-Rule Protocol Violations}

\stitle{\texttt{TEMP\_CENTER}.}
The clean reference uses trailing rolling operators.
In the implementation, \texttt{TEMP\_CENTER} is instantiated with \texttt{future\_days}=3: for each rolling statistic \(R_w[\cdot]\) in the affected feature constructor, the clean trailing operator is replaced by
\[
\widetilde{R}_w[x]_t = R_w[x_{\cdot+3}]_t,
\]
so the statistic indexed at day \(t\) depends on observations up to day \(t+3\).
Thus the chronological split is unchanged, but the effective feature information set reaches into the future.

\stitle{\texttt{NORM\_GLOBAL}.}
The clean reference fits feature standardization on the training split only.
If \((\mu,\sigma)\) denote the feature-wise mean and standard deviation, then clean normalization uses
\[
(\mu,\sigma) = \mathrm{Stats}(\mathcal{D}_{\mathrm{train}}),
\]
whereas \texttt{NORM\_GLOBAL} uses
\[
(\mu,\sigma) = \mathrm{Stats}(\mathcal{D}_{\mathrm{full}}),
\]
with \(\mathcal{D}_{\mathrm{full}}\) equal to the full panel passed to the benchmark runner.
This isolates preprocessing-scope leakage while leaving labels, execution, and portfolio construction unchanged.

\stitle{\texttt{STRUCT\_GRAPH}.}
The clean reference uses the trailing monthly graph \(W^{(m),\mathrm{trail}}\) defined above.
\texttt{STRUCT\_GRAPH} replaces it with a \emph{symmetric} 252-day graph centered on the first trading day of the month:
\[
\mathcal{W}^{\mathrm{sym}}_m
=
\{ s : \tau_m - 126 \le s \le \tau_m + 126 \},
\]
up to clipping at the beginning or end of the available date index.
The benchmark then computes the same absolute-correlation, top-\(k\), row-normalized peer matrix on \(\mathcal{W}^{\mathrm{sym}}_m\).
Hence the only rule change is that the graph estimator is allowed to use returns from both sides of the monthly anchor, including future returns relative to all dates \(t\) in month \(m\).

\stitle{\texttt{EXEC\_CLOSE}.}
The clean reference pairs close-of-day signals with next-day-open execution.
\texttt{EXEC\_CLOSE} instead uses same-day close entry and close-based realized returns:
\begin{align}
y^{\mathrm{exec\_close}}_{i,t}
&= \log \frac{C_{i,t+h}}{C_{i,t}}, \\
r^{\mathrm{trade,exec\_close}}_{i,t}
&= \frac{C_{i,t+1}}{C_{i,t}} - 1 .
\end{align}
This is important for implementation fidelity: the code changes \emph{both} the target and the one-day realized trading return to close-to-close semantics.
Therefore \texttt{EXEC\_CLOSE} is not ``close execution with an unchanged open-to-open label''; it is a fully close-based execution variant.

\stitle{\texttt{EXEC\_OPEN}.}
\texttt{EXEC\_OPEN} uses same-day open entry while still allowing the feature map to consume the full OHLCV bar of day \(t\), including post-open fields such as \(H_{i,t}, L_{i,t}, C_{i,t}\), and \(V_{i,t}\).
Its target and one-day realized trade return are
\begin{align}
y^{\mathrm{exec\_open}}_{i,t}
&= \log \frac{O_{i,t+h}}{O_{i,t}}, \\
r^{\mathrm{trade,exec\_open}}_{i,t}
&= \frac{O_{i,t+1}}{O_{i,t}} - 1 .
\end{align}
This is intentionally impossible under the clean decision-time semantics: the assumed fill time is the open of day \(t\), but the signal is allowed to depend on information revealed later in the same daily bar.

\paragraph{Why this supports one-rule attribution.}
Under this formalization, each switch modifies one benchmark dimension:
\texttt{TEMP\_CENTER} changes temporal feature support,
\texttt{NORM\_GLOBAL} changes normalization scope,
\texttt{STRUCT\_GRAPH} changes graph-estimation support,
\texttt{EXEC\_CLOSE} changes execution/label convention from next-open to same-close,
and \texttt{EXEC\_OPEN} changes execution timing from next-open to same-open while leaving the daily-bar feature constructor otherwise intact.
This is the sense in which the benchmark measures controlled protocol fragility rather than unconstrained pipeline drift.

\subsection{Metric Definitions}
\label{app:metric-definitions}

The main text uses a compact notation such as \(\mathrm{SR@5bps}\) and \(\mathrm{LG\mbox{-}SR@5bps}\).
We formalize these quantities here.

\paragraph{Per-day portfolio return with transaction costs.}
Let \(P_t(v;m,a,h)\) be the top-decile equal-weight portfolio selected on day \(t\) under protocol variant \(v\), market \(m\), model family \(a\), and horizon \(h\).
Let
\[
R^{\mathrm{gross}}_t(v;m,a,h)
=
\frac{1}{|P_t|}\sum_{i \in P_t} r^{\mathrm{trade}}_{i,t}(v;h)
\]
be the gross one-day portfolio return induced by the benchmark's execution convention.
Let \(\tau_t(v;m,a,h)\) be the benchmark turnover on day \(t\), and let \(c\) denote proportional transaction costs in basis points.
The net portfolio return used by the benchmark is
\[
R^{\mathrm{net}}_t(v;m,a,h,c)
=
R^{\mathrm{gross}}_t(v;m,a,h)
- \tau_t(v;m,a,h)\frac{c}{10^4}.
\]

\paragraph{Sharpe at \(c\) bps.}
Let \(\{R^{\mathrm{net}}_t(v;m,a,h,c)\}_{t=1}^T\) be the pooled daily net return series over all test windows under a fixed evaluation configuration.
Then the reported Sharpe ratio at \(c\) bps is
\[
\mathrm{SR@}c\mathrm{bps}(v;m,a,h)
=
\sqrt{252}\,
\frac{\mathbb{E}[R^{\mathrm{net}}_t(v;m,a,h,c)]}
{\mathrm{Std}[R^{\mathrm{net}}_t(v;m,a,h,c)]}.
\]
In particular, \(\mathrm{SR@5bps}\) is the case \(c=5\), and \(\mathrm{SR@10bps}\) is the case \(c=10\).

\paragraph{RankIC.}
For each evaluation date \(t\), let \(U_t\) denote the evaluated cross-section, let \(s_{i,t}\) be the model score for asset \(i \in U_t\), and let \(y_{i,t}^{(h)}\) be the realized horizon-\(h\) target.
The daily RankIC is the cross-sectional Spearman correlation
\[
\mathrm{RankIC}_t(v;m,a,h)
=
\rho_{\mathrm{Spearman}}
\big(
\{s_{i,t}\}_{i \in U_t},
\{y^{(h)}_{i,t}\}_{i \in U_t}
\big).
\]
The reported benchmark value is the mean over evaluation dates with a finite cross-sectional correlation:
\[
\mathrm{RankIC}(v;m,a,h)
=
\frac{1}{|T_{\mathrm{RankIC}}|}
\sum_{t \in T_{\mathrm{RankIC}}}
\mathrm{RankIC}_t(v;m,a,h).
\]

\paragraph{AUC.}
For each evaluation date \(t\), the binary target is defined by
\[
b_{i,t}^{(h)}=\mathbf{1}\{y_{i,t}^{(h)} > 0\}.
\]
The daily AUC is the cross-sectional ROC AUC of the model scores against these binary labels:
\[
\mathrm{AUC}_t(v;m,a,h)
=
\mathrm{ROC\mbox{-}AUC}
\big(
\{b_{i,t}^{(h)}\}_{i \in U_t},
\{s_{i,t}\}_{i \in U_t}
\big),
\]
with the standard rank-based interpretation in which ties receive half credit.
The reported benchmark value is the mean over evaluation dates that contain both binary classes:
\[
\mathrm{AUC}(v;m,a,h)
=
\frac{1}{|T_{\mathrm{AUC}}|}
\sum_{t \in T_{\mathrm{AUC}}}
\mathrm{AUC}_t(v;m,a,h).
\]

\paragraph{Leakage Gain in Sharpe.}
For any cost setting \(c\), the paired Leakage Gain in Sharpe is
\[
\mathrm{LG\mbox{-}SR@}c\mathrm{bps}(v;m,a,h)
=
\mathrm{SR@}c\mathrm{bps}(v;m,a,h)
-
\mathrm{SR@}c\mathrm{bps}(\mathrm{CLEAN};m,a,h).
\]
Thus \(\mathrm{LG\mbox{-}SR@5bps}\) is simply the variant-minus-clean difference in \(\mathrm{SR@5bps}\) under the same market, model family, horizon, and walk-forward grid.

\paragraph{Paired predictive metrics.}
The same paired construction is used for predictive metrics. In particular,
\begin{align}
\mathrm{LG\mbox{-}RankIC}(v;m,a,h)
&=
\mathrm{RankIC}(v;m,a,h)
-
\mathrm{RankIC}(\mathrm{CLEAN};m,a,h), \\
\mathrm{LG\mbox{-}AUC}(v;m,a,h)
&=
\mathrm{AUC}(v;m,a,h)
-
\mathrm{AUC}(\mathrm{CLEAN};m,a,h).
\end{align}

\paragraph{Generic paired metric notation.}
More generally, if \(q\) denotes any reported benchmark outcome, then
\[
\mathrm{LG}_{q}(v;m,a,h,c)
=
q(v;m,a,h,c)
-
q(\mathrm{CLEAN};m,a,h,c).
\]
The benchmark emphasizes \(\mathrm{LG\mbox{-}SR@5bps}\) in the main text because it combines predictive ranking, turnover, and transaction costs into one trading-level effect size.

\section{Implementation Details}
\label{app:implementation}

\subsection{Data Panels}
\label{app:data-panels}

Table~\ref{tab:data-characteristics} summarizes the two data panels and the clean protocol configuration.
Both panels are daily OHLCV equity panels with shared protocol semantics.
The U.S. panel is derived from S\&P 500 constituents and the China panel is derived from CSI 300 constituents.
The benchmark is constructed to compare protocol variants under matched OHLCV-level assumptions, not to provide a complete execution or capacity simulator.

\begin{table}[h]
\centering
\small
\caption{Dataset characteristics and clean protocol configuration.}
\label{tab:data-characteristics}
\begin{tabular}{lll}
\toprule
Property & US / S\&P 500 & China / CSI 300 \\
\midrule
Raw coverage & 2010-01-04 to 2024-12-30 & 2010-01-04 to 2024-12-31 \\
Unique tickers retained & 439 & 186 \\
Daily cross-section & 422 / 439 / 439 & 135 / 183 / 186 \\
(min / median / max) & & \\
Frequency & Daily OHLCV & Daily OHLCV \\
Test window & \multicolumn{2}{l}{2016--2024 yearly walk-forward} \\
Clean execution & \multicolumn{2}{l}{\(t+1\) open} \\
Labels & \multicolumn{2}{l}{Open-to-open returns} \\
Portfolio rule & \multicolumn{2}{l}{Top 10\% long-only, equal-weight} \\
Horizons & \multicolumn{2}{l}{\(h \in \{5,20\}\)} \\
\bottomrule
\end{tabular}
\end{table}

For each market, the universe, date range, preprocessing rules, and split definitions are fixed before evaluating protocol variants.
All clean and non-clean runs use the same retained tickers and the same yearly walk-forward test windows.
This ensures that Leakage Gain is not driven by changes in universe construction or test-period selection.

\subsection{Feature Schema and Label Construction}
\label{app:feature-schema}

All input features are derived from daily OHLCV bars.
The clean feature schema assigns every feature an admissibility timestamp.
At decision date \(t\), clean features may depend only on information available by the close of day \(t\).
The released feature-schema file enumerates each feature by name, source fields, transformation operator, lookback window, and admissibility timestamp.

Table~\ref{tab:feature-schema} summarizes the feature families and how they interact with the protocol variants.
The exact feature list and rolling windows are specified in the released schema/configuration files.

\begin{table*}[t]
\centering
\small
\caption{Feature-schema summary. The released schema file provides the exact feature names, source fields, rolling windows, and admissibility timestamps.}
\label{tab:feature-schema}

\setlength{\tabcolsep}{5pt}
\renewcommand{\arraystretch}{1.15}

\begin{tabularx}{\textwidth}{
  >{\raggedright\arraybackslash}p{0.18\textwidth}
  >{\raggedright\arraybackslash}X
  >{\raggedright\arraybackslash}X
}
\toprule
\textbf{Feature family} & \textbf{Clean construction} & \textbf{Related protocol variant} \\
\midrule

Raw OHLCV fields &
Daily open, high, low, close, and volume timestamped by bar availability &
\texttt{EXEC\_OPEN} tests use of post-open fields before the open \\

Return and momentum features &
Trailing returns computed from admissible historical bars &
Momentum uses fixed \texttt{ret\_20}; \texttt{TEMP\_CENTER} does not affect it \\

Rolling statistics &
Trailing rolling operators using observations no later than \(t\) &
\texttt{TEMP\_CENTER} replaces trailing windows with centered windows \\

Normalization &
Parameters fit within the training scope of each walk-forward window &
\texttt{NORM\_GLOBAL} fits normalization on the full sample \\

Relation / graph features &
Relations estimated only from admissible past data &
\texttt{STRUCT\_GRAPH} uses symmetric graphs with future returns \\

Labels &
Open-to-open returns aligned with the assumed entry price &
Execution variants change the timing consistency being tested \\

\bottomrule
\end{tabularx}
\end{table*}

For horizon \(h\), the clean label is an open-to-open \emph{log return} matched to the \(t+1\)-open entry convention.
For asset \(i\), if \(O_{i,t}\) denotes the open price on day \(t\), the clean horizon-\(h\) label is
\[
y^{(h)}_{i,t}
=
\log \frac{O_{i,t+1+h}}{O_{i,t+1}} .
\]
This matches the implementation in Appendix~\ref{app:formal-protocols}: the prediction target is a log return, whereas the one-day realized trading return used in the backtest remains a simple return.

\subsection{Walk-Forward Splits}
\label{app:walkforward}

All experiments use yearly walk-forward tests over 2016--2024.
For each test year, models are trained using only pre-test data available under the corresponding walk-forward split.
Any validation, early stopping, hyperparameter selection, or normalization fitting is restricted to pre-test data in the clean reference.
The same split definitions are reused across all protocol variants.

The split files released with the benchmark specify the exact train, validation, and test dates for each market and test year.
The paired design requires that the clean run and every protocol variant use the same split file.

\subsection{Model Families and Hyperparameters}
\label{app:model-details}

The benchmark uses two diagnostic baselines and four model-family extensions.
All model families are evaluated under the same markets, horizons, protocol variants, portfolio rule, and transaction-cost settings.
Hyperparameters are fixed across clean and non-clean variants; no protocol variant receives separate tuning.

\begin{table*}[t]
\centering
\small
\caption{Model-family summary. Hyperparameters are fixed across clean and non-clean variants.}
\label{tab:model-hyperparameters}

\setlength{\tabcolsep}{5pt}
\renewcommand{\arraystretch}{1.15}

\begin{tabularx}{\textwidth}{
  >{\raggedright\arraybackslash}p{0.15\textwidth}
  >{\raggedright\arraybackslash}p{0.23\textwidth}
  >{\raggedright\arraybackslash}X
}
\toprule
\textbf{Model family} & \textbf{Role in benchmark} & \textbf{Fixed hyperparameters / configuration} \\
\midrule

Momentum &
Fixed-rule diagnostic baseline &
Score \(=\) \texttt{ret\_20}; no trainable parameters \\

Ridge &
Linear trainable baseline &
Train-scope standardization; \(\ell_2\) penalty \(\alpha=1.0\) \\

LightGBM &
Tree-based non-linear model &
\texttt{num\_leaves=31}, \texttt{max\_depth=6}, \texttt{learning\_rate=0.05}, \texttt{n\_estimators=150}, \texttt{subsample=0.8}, \texttt{colsample\_bytree=0.8}, \texttt{reg\_alpha=0.1}, \texttt{reg\_lambda=0.1} \\

Graph MLP &
Graph-informed model &
Graph source \(=\) monthly top-5 return-correlation peer graph; base/relation branches \(32 \rightarrow 32\); head \(32 \rightarrow 16 \rightarrow 1\); optimizer Adam; learning rate \(10^{-3}\); batch size \(1024\); epochs \(1\) \\

LSTM &
Sequence model &
Lookback \(60\); hidden dim \(32\); layers \(1\); dropout \(0\); optimizer Adam; learning rate \(10^{-3}\); batch size \(512\); epochs \(1\); max train/val samples \(12000/3000\) \\

Transformer &
Attention-based sequence model &
Lookback \(60\); \(d_{\mathrm{model}}=32\); heads \(4\); layers \(1\); FFN dim \(64\); dropout \(0.1\); optimizer Adam; learning rate \(10^{-3}\); batch size \(512\); epochs \(1\); max train/val samples \(12000/3000\) \\

\bottomrule
\end{tabularx}

\vspace{2pt}
\begin{minipage}{0.96\textwidth}
\footnotesize
\emph{Note.} The hyperparameter configuration is shared across protocol variants. Thus, a clean run and its paired non-clean runs differ only in the selected protocol switch, not in model tuning, portfolio construction, transaction-cost assumptions, or test dates.
\end{minipage}
\end{table*}

The key reproducibility constraint is that the hyperparameter configuration is shared across protocol variants.
This ensures that measured Leakage Gains are attributable to the protocol switch rather than to retuning.

\subsection{Protocol Switches}
\label{app:protocol-switches}

Each benchmark run is specified by five identifiers:
\[
(\text{market}, \text{model family}, h, \text{transaction cost}, \text{protocol variant}).
\]
The clean reference and the five non-clean variants are implemented as explicit protocol switches.
Table~\ref{tab:protocol-switches} is a compact lookup table corresponding to the formal definitions in Appendix~\ref{app:formal-switches}.

\begin{table}[t]
\centering
\small
\caption{Protocol switches used in the benchmark.}
\label{tab:protocol-switches}

\setlength{\tabcolsep}{4pt}
\renewcommand{\arraystretch}{1.12}

\begin{tabularx}{\linewidth}{
  >{\raggedright\arraybackslash}p{0.18\linewidth}
  >{\raggedright\arraybackslash}p{0.26\linewidth}
  >{\raggedright\arraybackslash}X
}
\toprule
\textbf{Variant} & \textbf{Component changed} & \textbf{Intended diagnostic role} \\
\midrule

\texttt{CLEAN} &
None &
Decision-time-consistent reference \\

\texttt{TEMP\_CENTER} &
Rolling feature construction &
Temporal-causality leakage through future observations \\

\texttt{NORM\_GLOBAL} &
Normalization scope &
Global preprocessing leakage \\

\texttt{STRUCT\_GRAPH} &
Relation / graph construction &
Leakage through future-informed estimated structure \\

\texttt{EXEC\_CLOSE} &
Execution timing &
Same-day close execution inconsistency \\

\texttt{EXEC\_OPEN} &
Signal / fill alignment &
Use of post-open bar information while assuming same-day-open fills \\

\bottomrule
\end{tabularx}
\end{table}

For every paired comparison, the same market panel, model family, horizon, test dates, portfolio rule, and transaction-cost setting are used.
Only the protocol switch differs.

\subsection{Portfolio Construction and Metric Computation}
\label{app:portfolio-metrics}

On each evaluation date, stocks are sorted by the model score.
The top decile is held long-only with equal weights.
The exact per-day net-return convention and the formal definitions of \(\mathrm{SR@}c\mathrm{bps}\) and \(\mathrm{LG\mbox{-}SR@}c\mathrm{bps}\) are given in Appendix~\ref{app:metric-definitions}.
Here we only summarize the evaluation interface: all variants use the same top-decile long-only rule, equal weighting, and proportional transaction-cost accounting.
We report Sharpe ratios at 0, 5, and 10 bps, turnover, maximum drawdown, RankIC, AUC, and Leakage Gain.

For year-level stability, Leakage Gain is computed separately within each yearly test window.
Bootstrap confidence intervals are computed over yearly windows, and one-sided paired Wilcoxon signed-rank tests are applied to yearly paired gains for the dominant violations.

\subsection{Additional Cost-Sensitivity Check}
\label{app:cost-sensitivity}

\begin{figure*}[t]
  \centering
  \includegraphics[width=0.82\linewidth]{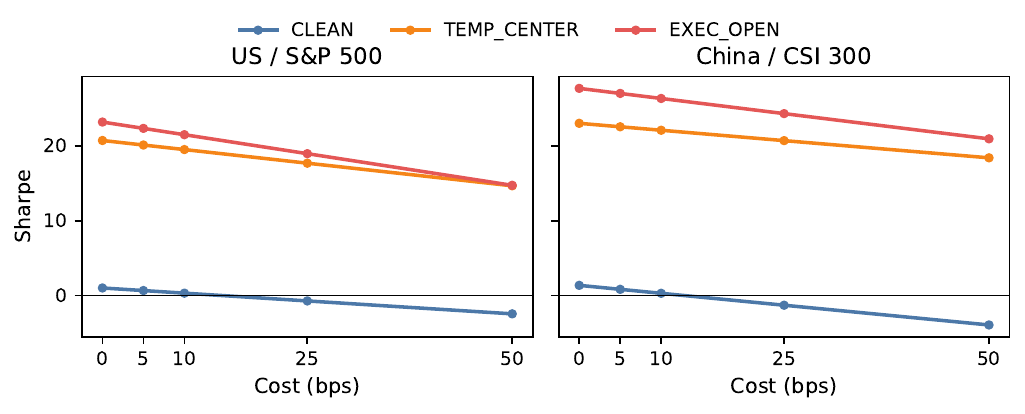}
  \caption{Transaction-cost sensitivity for the Ridge baseline at \(h=5\) on both markets.
  The clean Sharpe degrades toward or below zero as costs increase, but the two dominant violations remain strongly positive.
  The ordering between \texttt{TEMP\_CENTER}, \texttt{EXEC\_OPEN}, and \texttt{clean} is therefore not an artifact of fixing the evaluation at 5 bps.}
  \label{fig:appendix-cost-sensitivity}
\end{figure*}

To check whether the main ordering is specific to the 5 bps operating point, we reran the Ridge benchmark at \(h=5\) with transaction costs of \(0/5/10/25/50\) bps for \texttt{clean}, \texttt{TEMP\_CENTER}, and \texttt{EXEC\_OPEN}.
Figure~\ref{fig:appendix-cost-sensitivity} shows that the ranking remains stable on both markets throughout this wider range.
At 50 bps, the clean Sharpe falls to \(-2.43\) on the U.S. panel and \(-3.91\) on the China panel, whereas \texttt{EXEC\_OPEN} remains at \(14.73\) and \(20.93\), and \texttt{TEMP\_CENTER} remains at \(14.67\) and \(18.40\), respectively.




\section{Additional Year-Level Stability Results}
\label{app:year-stability}

\begin{table*}[t]
\centering
\footnotesize
\caption{Year-level stability results on the U.S. market (S\&P 500).}
\label{tab:year-stability-us}
\setlength{\tabcolsep}{5pt}
\renewcommand{\arraystretch}{1.12}
\resizebox{0.6\textwidth}{!}{
\begin{tabular}{@{}llp{4.15cm}p{4.15cm}@{}}
\toprule
\textbf{Model} & \textbf{$h$} & \textbf{TEMP\_CENTER} & \textbf{EXEC\_OPEN} \\
\midrule

Momentum
& 5  & \statcell{-0.00 [-0.00, 0.00]}{4/9, $p=0.250$}
     & \statcell{\textbf{6.57 [5.63, 7.61]}}{9/9, $p=0.002$} \\
& 20 & \statcell{-0.00 [-0.00, 0.00]}{4/9, $p=0.250$}
     & \statcell{\textbf{6.58 [5.63, 7.62]}}{9/9, $p=0.002$} \\
\addlinespace[1pt]

Ridge
& 5  & \statcell{25.20 [21.11, 29.29]}{9/9, $p=0.002$}
     & \statcell{\textbf{29.10 [24.00, 34.52]}}{9/9, $p=0.002$} \\
& 20 & \statcell{22.45 [18.69, 26.44]}{9/9, $p=0.002$}
     & \statcell{\textbf{23.50 [19.38, 28.14]}}{9/9, $p=0.002$} \\
\addlinespace[1pt]

LightGBM
& 5  & \statcell{25.24 [21.12, 29.42]}{9/9, $p=0.002$}
     & \statcell{\textbf{29.77 [24.53, 35.18]}}{9/9, $p=0.002$} \\
& 20 & \statcell{\textbf{21.88 [17.70, 26.20]}}{9/9, $p=0.002$}
     & \statcell{21.78 [17.10, 26.81]}{9/9, $p=0.002$} \\
\addlinespace[1pt]

Graph
& 5  & \statcell{23.96 [20.14, 27.82]}{9/9, $p=0.002$}
     & \statcell{\textbf{26.74 [22.31, 31.33]}}{9/9, $p=0.002$} \\
& 20 & \statcell{\textbf{20.58 [17.20, 24.17]}}{9/9, $p=0.002$}
     & \statcell{20.18 [15.91, 24.33]}{9/9, $p=0.002$} \\
\addlinespace[1pt]

LSTM
& 5  & \statcell{\textbf{6.22 [4.63, 7.72]}}{9/9, $p=0.002$}
     & \statcell{5.38 [2.70, 7.88]}{8/9, $p=0.006$} \\
& 20 & \statcell{\textbf{5.93 [3.94, 8.37]}}{9/9, $p=0.002$}
     & \statcell{5.65 [3.93, 7.56]}{9/9, $p=0.002$} \\
\addlinespace[1pt]

Transformer
& 5  & \statcell{\textbf{11.82 [8.87, 14.98]}}{9/9, $p=0.002$}
     & \statcell{7.29 [5.07, 9.77]}{9/9, $p=0.002$} \\
& 20 & \statcell{\textbf{7.29 [5.24, 9.17]}}{9/9, $p=0.002$}
     & \statcell{7.13 [4.74, 9.76]}{9/9, $p=0.002$} \\

\bottomrule
\end{tabular}
}
\vspace{2pt}
\begin{minipage}{0.92\textwidth}
\footnotesize
\emph{Note.} Each cell reports mean yearly $\mathrm{LG\mbox{-}SR@5bps}$ with the 95\% bootstrap confidence interval on the first line, and the number of positive years with the one-sided paired Wilcoxon $p$-value on the second line. Boldface marks the larger violation within each model--horizon pair.
\end{minipage}

\end{table*}
\begin{table*}[htbp]
\centering
\footnotesize
\caption{Year-level stability results on the China market (CSI 300).}
\label{tab:year-stability-cn}
\setlength{\tabcolsep}{5pt}
\renewcommand{\arraystretch}{1.12}
\resizebox{0.6\textwidth}{!}{
\begin{tabular}{@{}llp{4.15cm}p{4.15cm}@{}}
\toprule
\textbf{Model} & \textbf{$h$} & \textbf{TEMP\_CENTER} & \textbf{EXEC\_OPEN} \\
\midrule

Momentum
& 5  & \statcell{0.00 [0.00, 0.00]}{0/9, $p=\mathrm{NA}$}
     & \statcell{\textbf{6.58 [5.78, 7.47]}}{9/9, $p=0.002$} \\
& 20 & \statcell{0.00 [0.00, 0.00]}{0/9, $p=\mathrm{NA}$}
     & \statcell{\textbf{6.58 [5.78, 7.47]}}{9/9, $p=0.002$} \\
\addlinespace[1pt]

Ridge
& 5  & \statcell{23.18 [20.46, 26.17]}{9/9, $p=0.002$}
     & \statcell{\textbf{28.33 [24.83, 32.14]}}{9/9, $p=0.002$} \\
& 20 & \statcell{22.17 [19.32, 25.41]}{9/9, $p=0.002$}
     & \statcell{\textbf{23.80 [20.42, 27.55]}}{9/9, $p=0.002$} \\
\addlinespace[1pt]

LightGBM
& 5  & \statcell{23.18 [20.44, 26.25]}{9/9, $p=0.002$}
     & \statcell{\textbf{28.35 [24.46, 32.57]}}{9/9, $p=0.002$} \\
& 20 & \statcell{22.13 [18.79, 25.66]}{9/9, $p=0.002$}
     & \statcell{\textbf{24.18 [19.69, 28.84]}}{9/9, $p=0.002$} \\
\addlinespace[1pt]

Graph
& 5  & \statcell{22.40 [19.49, 25.65]}{9/9, $p=0.002$}
     & \statcell{\textbf{25.92 [22.25, 30.01]}}{9/9, $p=0.002$} \\
& 20 & \statcell{20.88 [17.71, 24.23]}{9/9, $p=0.002$}
     & \statcell{\textbf{21.24 [17.66, 24.84]}}{9/9, $p=0.002$} \\
\addlinespace[1pt]

LSTM
& 5  & \statcell{\textbf{11.52 [9.63, 13.60]}}{9/9, $p=0.002$}
     & \statcell{8.47 [5.96, 10.70]}{9/9, $p=0.002$} \\
& 20 & \statcell{7.86 [4.96, 10.46]}{9/9, $p=0.002$}
     & \statcell{\textbf{8.83 [6.67, 11.26]}}{9/9, $p=0.002$} \\
\addlinespace[1pt]

Transformer
& 5  & \statcell{\textbf{13.19 [9.46, 16.84]}}{9/9, $p=0.002$}
     & \statcell{11.56 [9.38, 14.05]}{9/9, $p=0.002$} \\
& 20 & \statcell{\textbf{14.22 [11.70, 17.37]}}{9/9, $p=0.002$}
     & \statcell{11.09 [7.65, 15.07]}{9/9, $p=0.002$} \\

\bottomrule
\end{tabular}
}
\vspace{2pt}
\begin{minipage}{0.92\textwidth}
\footnotesize
\emph{Note.} Each cell reports mean yearly $\mathrm{LG\mbox{-}SR@5bps}$ with the 95\% bootstrap confidence interval on the first line, and the number of positive years with the one-sided paired Wilcoxon $p$-value on the second line. Boldface marks the larger violation within each model--horizon pair.
\end{minipage}
\end{table*}


\end{document}